\documentclass[a4paper,10pt]{article}
\usepackage{graphicx}
\usepackage{latexsym,amsmath,verbatim}

\newcommand{\bea}{\begin{eqnarray}}
\newcommand{\eea}{\end{eqnarray}}

\title{Liquidity Crisis, Granularity of the Order Book and Price Fluctuations}

\author{M. Cristelli$^1$, V. Alfi$^{1,2}$, L.Pietronero$^{1,3}$, A. Zaccaria$^1$\\\\
$^1$ \small Dip. di Fisica, Universit\`a ``La Sapienza'', P.le A. Moro 2, 00185, Roma, Italy\\
$^2$ \small Centro ``E. Fermi'', Compendio Viminale, 00184, Roma, Italy\\
$^3$ \small  ISC-CNR, V. dei Taurini 19, 00185, Roma, Italy
}

\begin{document}

\maketitle

\begin{abstract}
We introduce a microscopic model for the dynamics of the order book to study
how the lack of liquidity influences price fluctuations. We use the average density of the stored orders (granularity $g$) as a proxy for liquidity. This leads to a Price Impact Surface which depends on both volume $\omega$ and $g$.  The dependence on the volume (averaged over the granularity) of the Price Impact Surface is found to be a concave power law function
$\langle\phi(\omega,g)\rangle_g\sim\omega^\delta$ with $\delta\approx 0.59$. Instead the dependence on the granularity is $\phi(\omega,g|\omega)\sim g^\alpha$
with $\alpha\approx-1$, showing a divergence of price fluctuations in the limit $g\rightarrow 0$. Moreover, even in intermediate situations of finite liquidity, this effect can be very large and it is a natural candidate for understanding the origin of large price fluctuations.
\end{abstract}

\section{Introduction}
One of the basic assumptions of the Arbitrage Pricing Theory, which is a standard tool of Financial Engineering, is the perfect liquidity of a market \cite{neocl,has,harris}.
This assumption implies that a market is always able to absorb incoming orders without 
producing a significant movement in the price.
However in the last ten years a great amount of studies demonstrates that finite liquidity effects play a crucial role in order to understand the features of a modern market \cite{farmer8,weber1,farmer6,farmer5,bouchaudmol,farbouch}.
From these studies liquidity appears more important in fixing the price response of a market than the order volume or the flow of information which are canonically assumed in Economics as the main ingredients to explain price movements~\cite{PietroGalle,Bouchaud6,marsiliok}. 
\\
The problem of finite liquidity have been highlighted by detailed studies of real order books  \cite{Bouchaud2,Bouchaud1,Bouchaud3}. 
In order to explain the empirical features of order books, several analytical and numerical models
have been introduced~\cite{slanina}.
A very important class of models for the order book are the so called zero-intelligence models which are characterized by random order deposition mechanisms \cite{challet2,farmer1}.
\\
In this paper we introduce and discuss a zero-intelligence model of order book to investigate the relation between finite liquidity and price fluctuations.
Zero-intelligence model are in general simple and workable but still able to 
reproduce the main properties of real order books.\\
In order to quantify the liquidity of a given configuration of the order book
we measure it via the average density of the stored orders in the book which 
we call {\itshape{granularity}}.
The analysis of the system response to a granularity fluctuation
can be properly addressed in the framework of a really elementary dynamics
for order placement which differs from the coarse-grained approach used in~\cite{farmer1,farmer2,farmer3}.
\\
We find that the impact of an order, which depends both on its volume and on the granularity of the configuration of the book, is a concave function of the volume.
This reproduces the empirical observations and represents an important test for the model itself. 
Our main result is that the price response to an incoming order is inversely proportional to the granularity. 
This means that granularity and the associated finite liquidity represent fundamental elements in the price response, which becomes divergent in the limit of vanishing granularity. 
From these results it is clear that a realistic modelling of a financial market requires the inclusion of finite liquidity.
\\
The paper is organized in the following way:\\
in section 2 we describe how an order book works and we introduce the problem of finite liquidity;\\
in section 3 we define our model and point out similarities and differences with respect to other order book models;\\
in section 4 we give the criteria to set the model parameters according to empirical data;\\
in section 5 we introduce granularity as an operational measure of liquidity;\\
in section 6 we introduce the Market Impact Surface and we study its dependence on both the volume and the granularity;\\
in section 7 we draw the conclusions and outline the possible extensions of our model.

\section{Order book and the role of liquidity}

\begin{figure}[!t]
\begin{center}
\vspace{1 cm}
\includegraphics[scale=0.35]{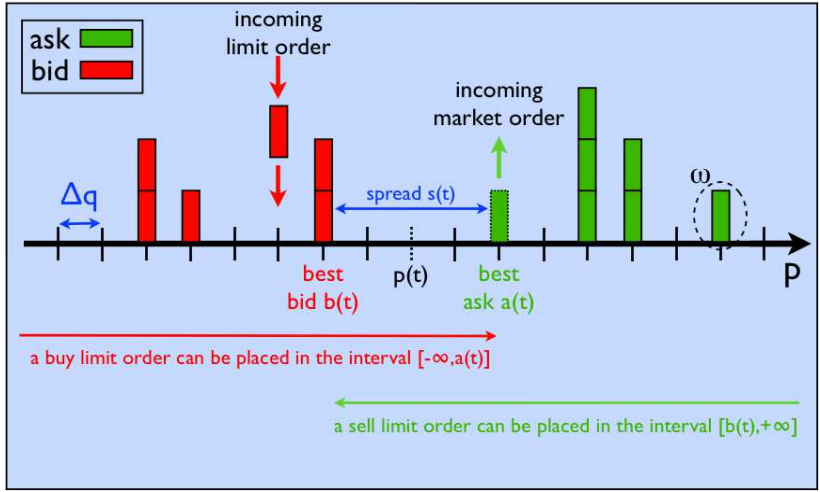}
\caption{Schematic representation of the order book dynamics. There are two classes of orders: the market orders and the limit orders. The market ones are orders of purchase or sale at the best available quote. On the other hand the limit orders are not immediately executed since they are placed at a quote which is less favorable than the best quote. The volume $\omega$ of the orders is an integer multiple of a certain amount of shares. The price of a stock can be defined as the mid-price $p(t)=[a(t)+b(t)]/2$.}
\label{ob2}
\end{center}
\end{figure}

\begin{figure}[!t]
\begin{center}
\vspace{1 cm}
\includegraphics[scale=0.35]{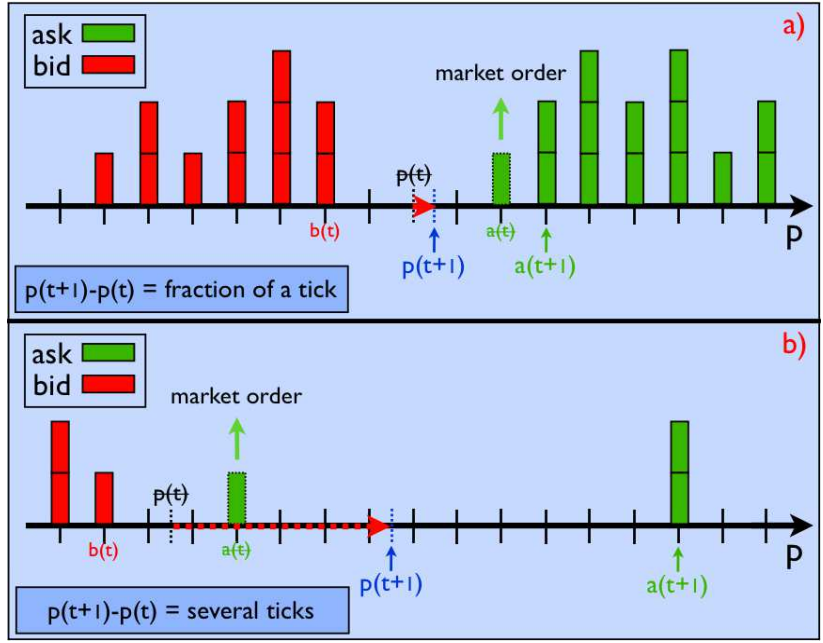}
\caption{A very liquid order book vs a liquidity crisis. This figure illustrates how the degree of liquidity of a market plays a key role in determining the system response to the volume of an incoming order. The order book is very liquid (\textit{panel a}) when a great amount of orders is stored in each side of the order book and almost all quotes behind the best one are occupied. In such a situation a market order produces a small perturbation of the system and then a small price adjustment.
On the contrary when a liquidity crisis occurs (\textit{panel b}), the order book is characterized by few orders stored and by a large average gap between adjacent orders. In this case even a market order with a small volume can produce a dramatic price fluctuation of several ticks. In our discussion high and low liquidity situations are symmetric and refer to the structure of the order book.}
\label{ob1}
\end{center}
\end{figure}

The elementary mechanism of price formation is a double auction where traders
submit orders to buy or sell. The complete list of all orders is called the 'order book'. In Fig. \ref{ob2} we give a schematic representation of an order book.
There are two classes of orders: market orders and limit orders. 
Market orders correspond to the intention of immediately purchase or sale at the best price
(quote) available at that time.
The limit ones instead are not immediately executed since they are offers to buy or sell at a certain quote which is 
not necessary the best one. If we consider a sell limit order this means that its quote is higher than (or equal to) the best bid $b(t)$ which is the order of buying with the highest price. On the other hand 
a buy limit order implies that its price is lower than (or equal to) the best ask $a(t)$ which is the order of selling with the lowest price.
The non-zero difference between $a(t)$ and $b(t)$ is defined as the spread $s(t)=a(t)-b(t)$. 
The prices of placement of orders (called 'quotes') are not continuous but quantized in unit of ticks
whose size is an important parameter of an order book. 
\\
The price of a stock can be conventionally defined as the mid-price $p(t)=[a(t)+b(t)]/2$ and it can change only if a limit order falls inside the spread or if a market order matches all the orders placed at the best quote. 
It is clear that the specific configuration of an order book it is a very important
aspect for price movements. A thick book full of orders can absorb the arrival of new orders
without giving rise to large jumps of the price. On the contrary if the book is sparse, even a small
incoming order could trigger a large price variation.
In order to clarify this problem we can consider an order book in the configuration of \textit{panel a} in Fig. \ref{ob1}. 
This situation corresponds to a very liquid market in which the order book can absorb several orders without 
large price variations even if they are relatively large. This regime corresponds to the assumptions of the standard financial picture where an order can be immediately executed, its impact is marginal and the market is nearly efficient.
\\
Instead if a liquidity crisis occurs, the configuration of the order book changes dramatically and the situation is like the one represented in \textit{panel b} in Fig. \ref{ob1}. The orders stored are few, the average distance between them is large and the flow of market orders cannot be easily and immediately absorbed by the system and by consequence
even a small order can produce a large price variation. 
Hence a lack of liquidity can dramatically amplify the response of the system.\\

\section{Definition of the model}\label{sec:2}

The purpose of this paper is to study the price response function in presence of liquidity crisis and to
appropriately address this problem we are going to introduce a model with a suitable
microscopic dynamics. Our model can be defined as a 'zero-intelligence' model as the ones introduced in \cite{farmer1,farmer2}. 
It should be noted that the specific questions that we consider cannot be addressed by the dynamics of Refs.  \cite{farmer1,farmer2}.\\
In fact in \cite{farmer1,farmer2} the deposition is modelled in term of a flow of orders driven by a Poisson process. The orders arrive at each quote at a certain fixed rate so that the deposition interval and the number of orders per unit of time are infinite but the orders per unit price interval are finite. The motivation of the authors of \cite{farmer1,farmer2} is that in such a way the order book can be never depleted and some finite size effects, due to a finite deposition interval, are avoided. \\
On the other hand in our model it is preferred an elementary \textit{microscopic} mechanism for order deposition. 
\\
The difference between these two mechanisms for order deposition is
the definition of the time unit.
In the present paper the time unit corresponds to the time (that is not fixed) between two following orders. Instead in Refs. \cite{farmer1,farmer2} the time unit corresponds to a physical and fixed amount of time (for example $\Delta t$ equal to some minutes) during which several actions are made 
by investors. This permits a \textit{coarse-grained} description in terms of average quantities, such as order flow. 
\\
For the specific implementation of our model we start by considering three mechanisms of deposition which will be compared with the empirical data. \\ We want to point out that only one of the three mechanisms gives rise to a realistic dynamics.
Certain elements are however common to the three mechanisms and we start by discussing them.
\\
In our set-up, as in a real order book, an order can be placed at a quote that is an integer multiple (positive or negative) of the tick size $\Delta q$. In this paper we do not investigate the effects of this parameter; therefore we set it equal to $1$. 
For the sake of simplicity we assume that all orders (limit and market) have the same volume $\omega=1$. The fact that a market order has the same volume of a limit order is coherent with the observation that in more than $99\%$ of the cases a market order does not exceed the best limit order available \cite{farmer6}.
\\
In this model there is one mechanism for order creation, the limit orders deposition, and two for order annihilation, the arrival of market orders and the cancellation process of limit orders. The cancellation of a limit order occurs when the order has been stored in the book for $\tau$ time steps without being executed, where $\tau$ is a parameter to be fixed\footnote{We know that this assumption, reasonable for the present study, 
is far to be realistic from the moment that 
analysis of real order books have shown that the lifetime of a limit order increases monotonically with its distance from the mid-price~\cite{Bouchaud1,farmer3}}. 
The balancing of the three mechanisms of creation/annihilation fixes the properties of the steady state that is reached very quickly (about $5\times10^3$ steps if $\tau$ ranges between $100$ and $1000$). 
Two limit orders can be also removed from the order book when the spread $s(t)$ is zero, that is when an incoming limit order is placed at the opposite best quote. However, these events are very unlikely ($<1\%$ of the number of the market orders) and their effect is negligible.
\\
We define the price as the mid-price $p(t)=[a(t)+b(t)]/2$ where $a(t)$ and $b(t)$ are the quotes of the best ask and of the best bid. 
\\
In our model we completely neglect the daily closures of the market and therefore the strong price variations during the night.
The typical length of a run is $10^6$ time steps ($10^7$ if a larger sample is needed) which corresponds to a real sample of about $5$ years, estimating about $10^3$ operations per day.

\subsection{Order deposition}\label{depo}
 
Now we are going to discuss three possible cases for order deposition which we will test with respect to empirical data in order to select the most appropriate mechanism. 
Due to the symmetry of the order book,
the probability that an order is a sell or a buy one is always the same.
\subsubsection{Case 1}
Once the nature (buy or sell) is determined, the order can be a market one with probability $\pi$ and a limit one with probability $1-\pi$, with $\pi<0.5$. Limit orders will be placed in the interval $[b(t),b(t)+L]$ or $[a(t)-L,a(t)]$ depending on their nature (sell or buy respectively).
The probability distribution within these intervals is considered uniform and $L$ is a free parameter to be fixed.
Hence the deposition interval is independent on the spread size.  
\subsubsection{Case 2}
The order can be a market one with probability $\pi$ and a limit one with probability $1-\pi$ with $\pi<0.5$ as in case $1$. A limit order will be placed in the interval $[b(t)+1,b(t)+k\,s(t)]$ or $[a(t)-k\,s(t),a(t)-1]$ with $k>1$  with a uniform distribution, being $s(t)$ the spread. In this case the length of the deposition interval depends on the spread size through a sort of auto-regressive process. This implies that the probability that a limit order produces a price variation (i.e. an order is placed inside the spread) is time-independent and approximately equal to $k^{-1}$. We anticipate that this second case will turn out to be the more realistic one with respect to the empirical data.
\subsubsection{Case 3}
The third case is inspired by the mechanism proposed in \cite{farmer3}. First a number $\xi$ in the interval $[-L,L]$ is extracted. If $\xi<s(t)$ the order is a limit order otherwise it is a market one. If the order is a sell limit order its quote is $a(t)-\xi$. On the contrary if the order is a buy one its quote is $b(t)+\xi$. In such a way the probability of being a market or a limit order is not set a priori.  
\\
\\
We have previously mentioned that our choice for order deposition produces a finite number of orders stored in the order book and a finite depth. This causes a finite size effect: the order book is subject to complete depletion with a cut off in those quantities linked with the book depth. However the probability of a complete depletion goes very quickly to $0$ when the cancellation parameter $\tau$ is increased (this probability is less than $10^{-7}$ for $\tau>200$).\\
Moreover we note that the uniform deposition mechanisms used in this paper cannot reproduce in a realistic way all the empirical features of the order book; nevertheless this assumption has been made with the aim to define a minimal model still able to reproduce the main properties of the order book. Besides, the main goal of this work is the evaluation the effects of the granularity in price fluctuations, which depends on the static realization of the profile.

\section{Calibration of parameters and preliminary results}\label{sec:3}

\begin{figure}[htbp]
\begin{center}
\vspace{1 cm}
\includegraphics[scale=0.215]{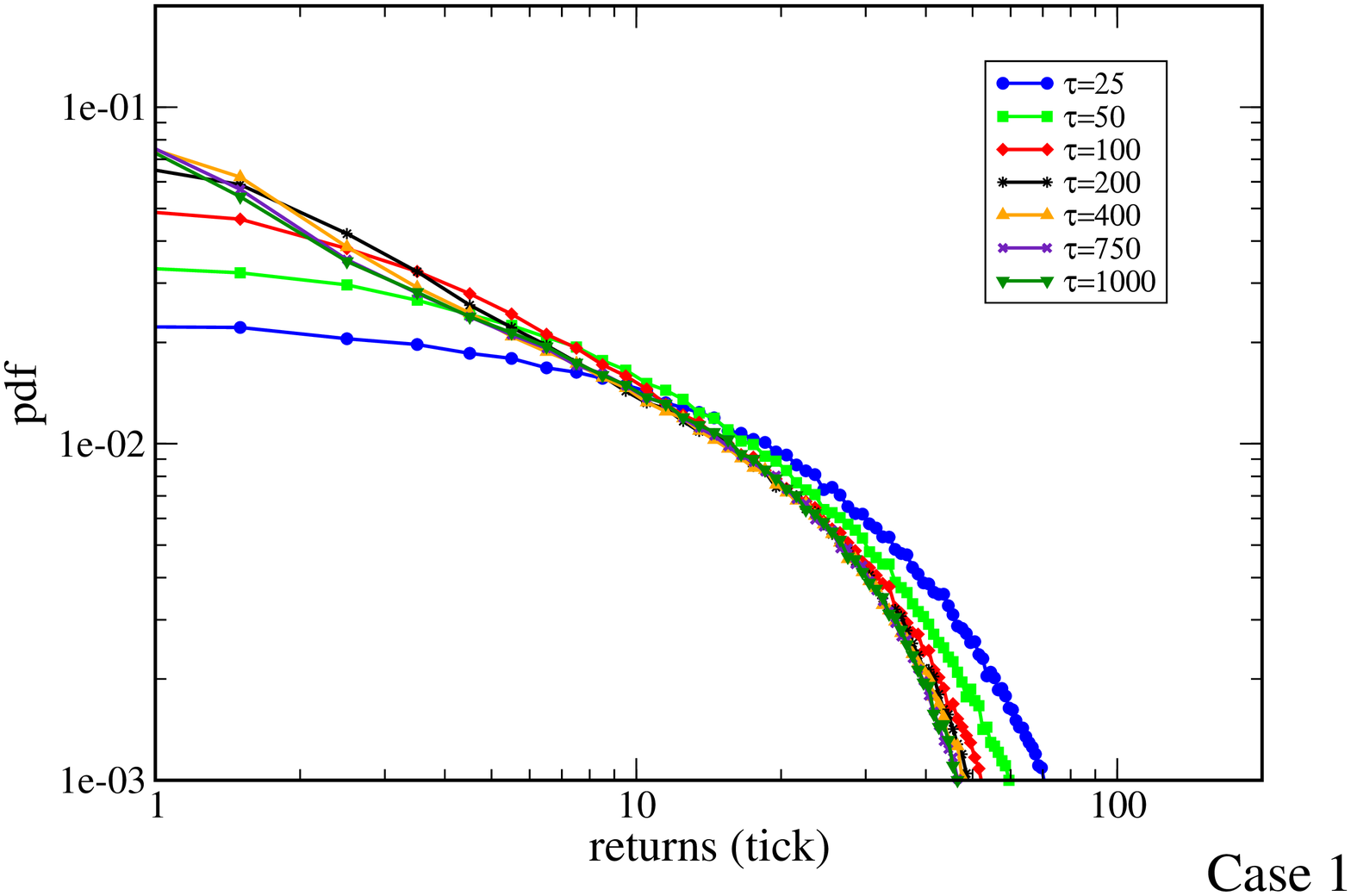}
\includegraphics[scale=0.215]{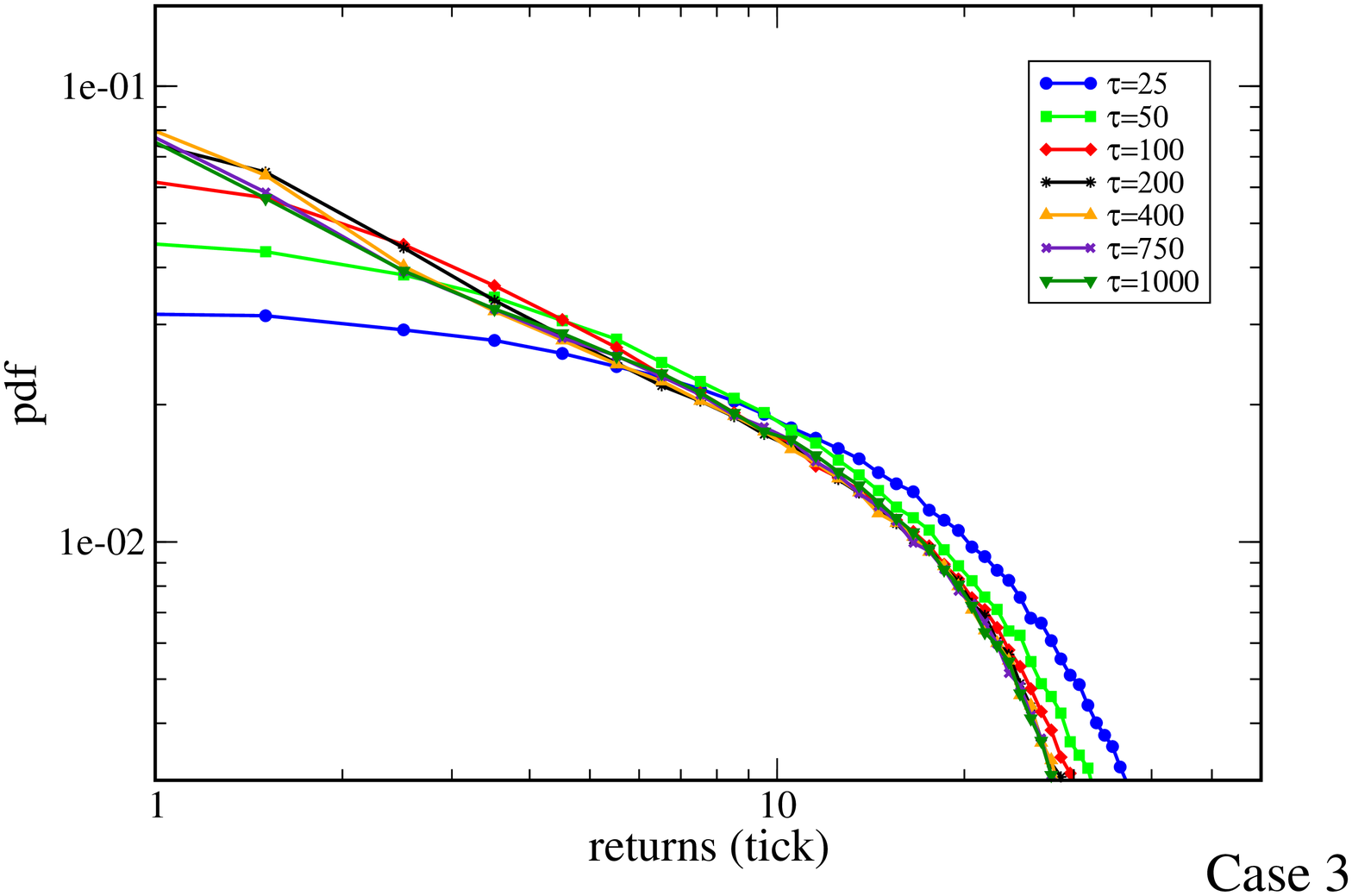}

\includegraphics[scale=0.3]{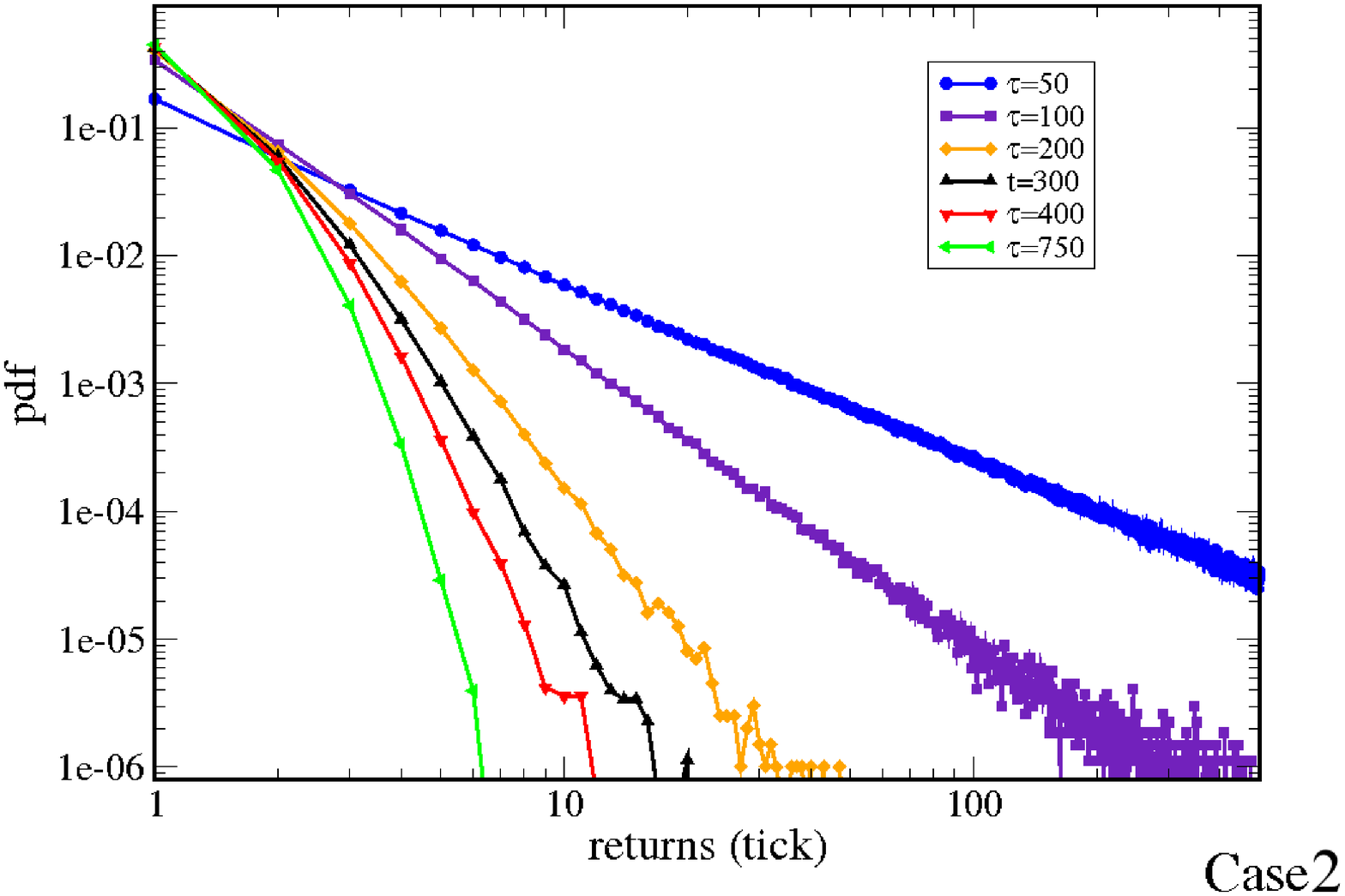}
\caption{Fat Tails. We show the probability density function of returns for the three mechanisms for order deposition and for different values of $\tau$. Because of the symmetry of the model the left tail is identical to the right one. We see that all three mechanisms give rise to a power law behavior but only the case 2 produces realistic values for the tail exponent $\gamma$: $\gamma=-3.1$ for $\tau= 200$ and $\gamma=-5.4$ for $\tau=400$. In fact for cases $1$ and $3$ the exponent $\gamma$ is larger than $-2$ which is not realistic. Moreover we note that the case 1 and 3 are very similar with respect to the tail exponent and to the cut off around $100$ due to the deposition mechanism. We conclude that these two mechanisms produce the same results except for slight differences. Conversely the \textit{auto-regressive} deposition mechanism of the case 2 appears to belong to a completely different class. This fact is also supported by the fact that the case 2 reproduces a certain volatility clustering differently from case 1 and 3.}
\label{pdfpdf}
\end{center}
\end{figure}

In this section we check the properties of the models we have introduced with respect to the empirical data. This analysis will also allow us to decide how to calibrate the parameters. 
\\
The empirical results of Ref. \cite{farmer6} indicate that a realistic value for $\pi$ is approximately $1/3$. In case $3$ the probability of being a market or a limit order is not fixed a priori. Nevertheless in this case we find that the fraction of market orders tends to a realistic value (about $0.32$) when $\tau$ grows, and becomes substantially independent on $\tau$ when the probability of emptying the order book turns to be negligible (i.e. $\tau>200$). This feature does not seem to depend crucially on the parameter $L$.
\\
We now have to fix the remaining parameters. 
Two quantities that can be calibrated on empirical data are the average spread and the average number of stored orders $n$. The parameters of the model which rule these quantities are 
$\tau$, $k$ (case $2$) and $L$ (case $1$ and $3$).
Since it is not always possible to a find a configuration of all parameters able to 
reproduces these two quantities in a realistic way, we give priority to those sets of parameters that produce a reasonable average number of orders. 
\\
All three cases exhibit a realistic accumulation of orders (i.e. $50-100$ orders per side) for $\tau$ ranging from $200$ to $500$. This clarifies the role of $\tau$ in driving the average number of orders in the steady state. In case $1$ we choose $L=200$ and in case $3$ $L=100$, in such a way the order book has approximately the same depth in both cases. For these two cases we applied the previous criterion of priority in order to have a realistic average number of orders. 
Case $2$ appears to be very interesting because it exists a range of parameters for which we can reproduce both a realistic average spread and average number of orders. For instance for $\tau=400$, $k=4$ we obtain an average number of orders per side of about $60$ and $\langle s \rangle\approx2.8$ ticks. In Table \ref{para} we report a summary of the results for the tested parameters. We have only reported some selected results and a more detailed and systematic analysis will be given in a future survey \cite{futob}.\\
\begin{table}[!htbp]
\begin{center}
\begin{tabular}{|c||c|c|c|}
\hline
 &Case 1& Case 2 & Case 3\\ \hline\hline
$\pi$&$1/3$&$1/3$&/\\ \hline
 &$\tau<200$&$\tau<300$&$\tau<200$\\ 
&$n\ll50$&$n\ll50$& $n\ll50$\\
& & &\\
$\tau$&$200<\tau<500$&$300<\tau<750$&$200<\tau<500$\\
& $50<n<100$&$50<n<100$&$50<n<100$\\
& & &\\
&$\tau>500$ &$\tau>750$&$\tau>500$\\
&$n\gg100$ &$n\gg100$ &$n\gg100$ \\
\hline
$L$&200&/&100\\ \hline
$k$&/&$3$ or $4$&/\\ \hline
General&&Realistic case&\\
remarks&$\langle s \rangle \gg 20$&$\langle s \rangle=2.8$ &$\langle s \rangle \gg 20$\\ 
&&for $k=4$, $\tau=400$&\\ \hline
\end{tabular}
\end{center}
\caption{Summary of the setting of the parameters. Only case 2 is able to reproduce reasonable values for the quantities of interest.}
\label{para}
\end{table}
\\\\
It is worth noticing that the \textit{auto-regressive} case $2$ is also interesting with respect to the Stylized Facts. In fact only in this case we can observe some volatility clustering in the price increments (returns). All cases exhibit a return probability density function which decades as a power law (see Fig. \ref{pdfpdf}) but only in the second case the exponent $\gamma$ of the tail is compatible with the empirical observations ($-5<\gamma<-2$). For instance in case $2$ we find $\gamma=-4.1$ for $\tau= 200$ and $\gamma=-5.4$ for $\tau=400$. 

\section{Operational estimator of the granularity}\label{sec:4}
\begin{figure}[!ht]
\begin{center}
\vspace{1cm}
\includegraphics[scale=0.40]{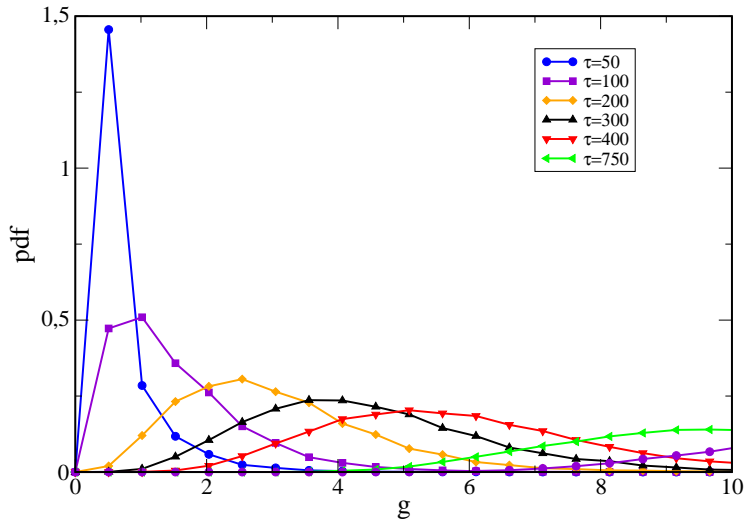}
\caption{Probability density function of the granularity $g$ for the case $2$ for different values of $\tau$. Obviously large values of $\tau$ allow for a significant accumulation of orders at every quote (i.e. $g>1$), despite this fact the probability of $g<1$ and consequently of large price variation is not negligible for $\tau\approx400$ which has been previously recognized as a suitable choice of this parameter. }
\label{pdfg}
\end{center}
\end{figure}

Given a certain configuration of the order book we want to define a way to measure the liquidity of it. In fact our objective is a detailed study of the relation between finite liquidity and price fluctuations.
\\
We would like to describe the effect that the order book may not be compact but characterized by orders separated by voids.
In order to perform this analysis we are going to introduce a function $g$ which we call granularity.
We propose a definition for granularity which is related to the inverse of the size of the average void between two adjacent quotes (gaps). At each time $t$ the spread $s(t)$ sets a characteristic length which can be used as unit of measurement for these gaps.
However we are also interested in measuring the granularity of the order book in a region far from the best quotes since the two sides of the order book usually have a depth much larger than $s(t)$. Therefore we define a partitioning of size $s(t)$ of one side of the order book and we perform the following average
\begin{equation}
\langle g(t)\rangle=\frac{1}{N(t)}\sum_{j=1}^{N(t)}\frac{n_{s(t),j}}{s(t)} 
\end{equation}
where $N(t)$ is the number of intervals of length $s(t)$ defined by the partition (we require $N(t)>2$ in order to calculate $g(t)$) and $n_{s(t),j}$ is the volume of orders in an interval of length $s(t)$. From the definition of $N(t)$ we observe that $N(t)s(t)=\bar{L}$ is approximately the range of the order book and consequently
the granularity $g$ results equal to the average linear density of orders
\begin{equation}\label{gg}
\langle g(t) \rangle =\frac{\Omega}{\bar{L}}
\end{equation}
where $\Omega=\sum_{j=1}^{N(t)}n_{s(t),j}$. For the sake of simplicity we now drop the temporal dependence of $g$ and the brackets of the average. We note that the definition of granularity given in Eq. (\ref{gg}) does not measure the real average size of the voids because more than one order can be stored at the same quote. Therefore $g^{-1}$ can be seen as the equivalent gap between two adjacent quotes in an hypothetical system where each quote can store only one order.
We also observe that in the limit of a continuous order book, the price variation $\Delta p$ produced by an order
of volume $\omega$ would be
\begin{equation}
\int_{0}^{\Delta p} \rho (p) dp = \omega
\end{equation}
where $\rho(p)$ is the density of the stored orders.
If we approximate $\rho(p)$ with its average value $g$ we find the following scaling relation for $g$ and $\omega$
\begin{equation}\label{scalinggw}
\Delta p = \frac{\omega}{g}.
\end{equation}
In this framework a liquidity crisis occurs when $g\rightarrow0$, on the contrary the market is very liquid if $g\gg1$. In fact when $g$ is very large a great amount of orders can be executed without producing a significant price variation.
\\
The granularity defined by Eq. (\ref{gg}) has the dimension of a $price^{-1}$. In order to obtain an absolute parameter for liquidity, we define the dimensionless liquidity as $\bar{g}=g\Delta q$. As we have seen in section \ref{sec:2} at this stage we are not interested in the effect of the tick size $\Delta q$ on the model so we have set it equal to $1$. Consequently  the dimensional and the dimensionless liquidity numerically coincide. Furthermore we want to point out that the definition of $g$ given by Eq. (\ref{gg}) is not directly related to a specific disposition in the order book but it gives only an average information about the void size.
\\
In Fig. \ref{pdfg} we show the histogram of $g$ for the model labelled as 'case $2$'.
The probability to observe a liquidity crisis, as expected, vanishes when $\tau\rightarrow\infty$ because of the large amount of orders stored in the order book. However, this probability cannot be neglected for values of $\tau$ ranging from $50$ to $400$ that we have previously recognized as the realistic ones.
\\
It is important to note that in this paper we give a different definition of granularity with respect to the definition of Ref.  \cite{farmer2}. In fact in that paper the authors define two dimensionless parameters to describe the discreteness of the system: the first, named granularity, is the dimensionless order size, the second one is the dimensionless tick size. In a certain sense both aspects are included in our definition of granularity and combined in a single parameter. These different choices are motivated by different purposes, while the authors of \cite{farmer2} gives the priority to the parameters that describe the order flows, differently we are going to focus our attention on how the discreteness influences the response function of the system.
\section{Price Impact Surface}\label{sec:5}

In this section we restrict our analysis to the 'case $2$'  mechanism for order deposition and all the results reported here correspond to $\tau=400$ and $k=4$. 
In order to analyze the average response of the system to an external perturbation (i.e. the volume of an incoming market order) after a certain time lag $\Delta t$, we introduce the Price Impact Function $\eta$ defined as
\begin{equation}\label{impact1}
\eta(\omega, \Delta t)= 2 E[\Delta p(\Delta t)|\omega].
\end{equation}
One can consider two interesting limits of Eq. (\ref{impact1}), the asymptotic limit ($\Delta t\rightarrow\infty$) and the instantaneous limit ($\Delta t\rightarrow0$). In this paper we restrict our investigation to the second case because the persistent effects on the market cannot be fully explained by a zero-intelligence model which neglects all the time-correlated structures of the order deposition (see\cite{farmer5,bouchaudmol,farbouch,Bouchaud5,farmer4}).\\
Now we are going to consider also 
the dependence of $\phi$ on the granularity $g$.
In this way we have to deal with a Price Impact \textit{Surface} rather than a Price Impact Function and we can
analyze how the price response of the system depends both on the order volume and on the granularity $g$.
Hence we define the Price Impact Surface as
\begin{equation}\label{impact2}
\phi(\omega, g)= 2 E[\Delta p|\omega,g].
\end{equation}
Before investigating the dependence of the price response surface on $g$ we should verify the well-known empirical property that $\langle\phi(\omega, g)\rangle_g$ (i.e. the standard Price Impact Function defined as in Eq. (\ref{impact1})) is a concave function with respect to the order volume $\omega$. This is a manifestation of the fact that the order book is far from being in a linear regime. While the agreement on the concavity is almost universal, different functional form of the Impact Function have been given: the authors of \cite{Bouchaud1} propose $\phi(\omega)\sim \ln \omega $, instead other authors propose $\phi(\omega)\sim \omega^\delta $ with exponent $\delta\approx0.5$ (\cite{stanley1}) and $\delta\approx0.3$  (\cite{farmer5,farmer7}). In Fig. \ref{surfacew} we find that the surface averaged on $g$ (black crosses) is concave and follows a power law 
\begin{equation}
\langle\phi(\omega, g)\rangle_g\sim\omega^\delta\,\,\,\text{with}\,\,\,\delta\approx 0.59.
\end{equation}

\subsection{Price Impact Surface in the direction of $\omega$}\label{sec:51}

\begin{figure}[!t]
\begin{center}
\vspace{1cm}
\includegraphics[scale=0.40]{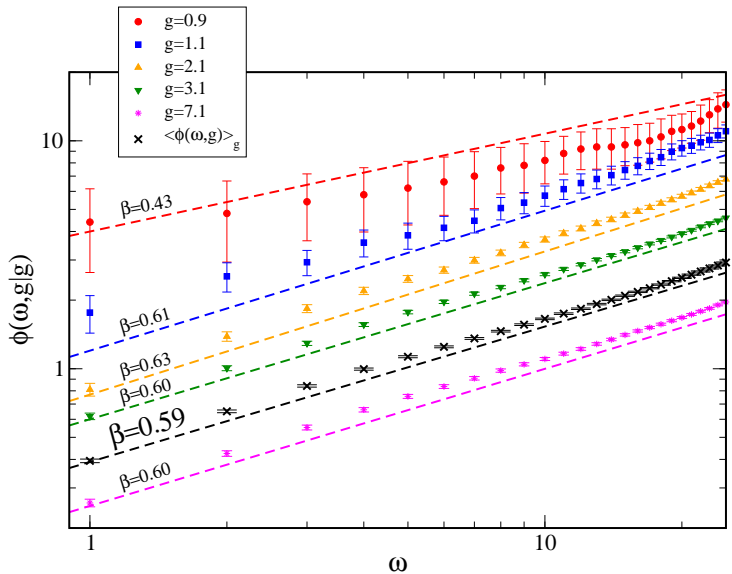}
\caption{Market Impact Surface in the direction of volumes $\omega$ for case $2$ and $\tau=400$. We can see that the function is concave as the one observed for real data. Its behavior is approximately a power law $\omega^\beta$. The dashed lines correspond to the results of the fitting procedure and they are shifted for clarity purposes. The price response for large values of $g$ cannot be larger than the one for small values of $g$. The black crosses represent the impact function averaged on $g$ and correspond approximately to an exponent of $0.59$. }
\label{surfacew}
\end{center}
\end{figure}

Now we consider the Price Impact Surface as a function of $\omega$ for fixed values of $g$. We see from Fig. \ref{surfacew} that its behavior is a power law where the exponent $\beta$ and the amplitude depend on $g$
\begin{equation}
\phi(\omega,g|g)\sim\omega^\beta.
\end{equation}
This function is highly concave for small value of $g$ and $\beta$ grows for increasing values of $g$. However we notice that the dependence of the amplitude on $g$ is far stronger than the one of $\beta$. This implies, as expected, that the Impact Surface produces smaller variation of price in the limit of large $g$, even if the exponent $\beta$ is larger.
\\
It is worth noticing that the observed scaling behavior for $\omega$ is different from the one predicted by the zero order approximation made in Eq. \ref{scalinggw}. This is due to the peculiar shape of the order book profile whose maximum is peaked far from the best quote. 
\\
The results of Fig. \ref{surfacew} are qualitatively similar to those of Fig. 6 of Ref. \cite{farmer2} even if the different definitions of granularity do not permit a detailed comparison of the two models. 

\subsection{Price Impact Surface in the direction of $g$}\label{sec:52}

\begin{figure}[!t]
\begin{center}
\vspace{1cm}
\includegraphics[scale=0.40]{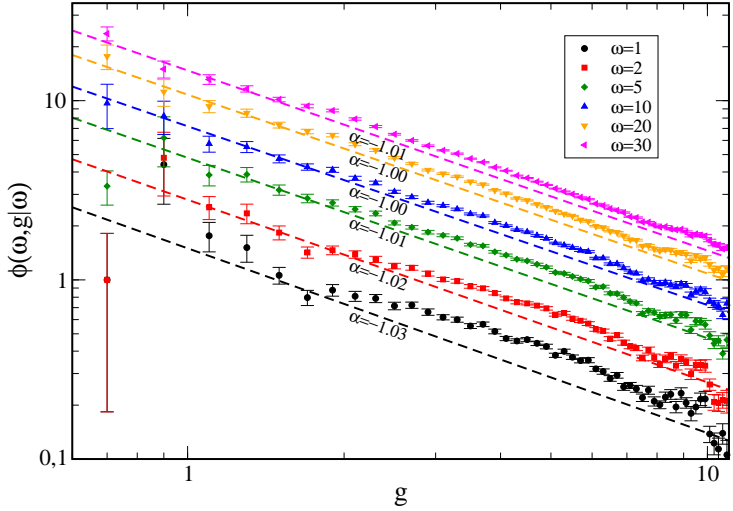}
\caption{Market Impact Surface in the direction of the liquidity $g$ for case $2$ and $\tau=400$. In this direction the surface can be fitted with a power law $g^\alpha$ with a negative exponent. This implies that when $g$ tends to $0$ (i.e. when a liquidity crisis occurs) the price response can be very large even if the external perturbation (i.e. the volume $\omega$ of an incoming order) is very small. The exponent $\alpha$ is approximately independent on $\omega$ and its value is compatible with $-1$ within error bars.}
\label{surfaceg}
\end{center}
\end{figure}

Now we analyze the price variation as a function of the granularity $g$ for a fixed volume $\omega$ of the incoming market order. We find that the $g$-dependence of the surface is well represented by a power law 
\begin{equation}
\phi(\omega,g|\omega)\sim g^\alpha\,\,\,\text{with}\,\,\,\alpha\approx-1.
\end{equation}
In this case the exponent is negative and it appears to be nearly independent on $\omega$ (see Fig. \ref{surfaceg}). In fact we find that its value is $-1$ (within the error bars) for all values of $\omega$. Consequently when the granularity vanishes the response of the system diverges. We can now quantitatively analyze how this limit is approached for finite
(but large) values of $g$. For example if we consider an order of size $\omega=10$, 
the average price variation induced when $g\approx0.5$ is about ten times greater than the one observed when $g\gg1$.
\\
Therefore we have found that the amplification induced by the discreteness of the system is approximately proportional to the equivalent gap measured by $g$ as predicted by the zero order approximation of section \ref{sec:4}.
However we want to stress that for small values of $g$ some deviations from the $g^{-1}$ scaling behavior seem to be observed. This fact suggests that, when $g$ vanishes, the deviation of 
the density $\rho(p)$ from the average density measured by $g$, is significant.
Further investigation of this aspect will be considered in future works.
\\
The fact that the exponent of the power law in the direction of $g$ does not seem to exhibit a significant dependence on $\omega$ may suggest a possible factorization of the price impact surface, as we are going to verify in the next section.

\subsection{Factorization of the Price Surface Impact}\label{sec:53}
\begin{figure}[htbp]
\begin{center}
\vspace{1cm}
\includegraphics[scale=0.40]{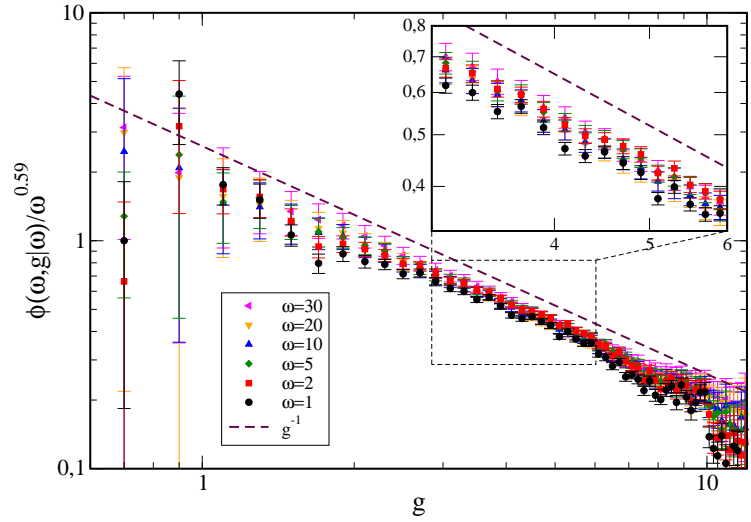}
\caption{Factorization of the Price Impact Surface. In order to verify if the Price Impact Surface admits the factorization $\phi(\omega,g)=\varphi(\omega)\psi(g)$, we rescale the surface in the direction of g (i.e. $\phi(\omega,g|\omega)$ ) with a function of $\omega$. For the sake of simplicity we choose the Impact Function averaged on $g$ that we have previously found to be proportional to $\omega^{0.59}$. We obtain a very good collapse of the rescaled functions within error bars. This confirms that the Price Impact Surface can be factorized at least for small values of $\omega$. In the insert we show a magnification of the rescaled functions to highlight their statistical compatibility.}
\label{collapse}
\end{center}
\end{figure}

In Ref. \cite{Bouchaud1} the authors observe that the empirical impact function admits the following factorization
\begin{equation}
\eta(\omega,\Delta t)=f(\omega)h(\Delta t).
\end{equation}
Similarly we verify whether the Impact Surface Function could be expressed in the following way
\begin{equation}\label{complete}
\phi(\omega,g)=\varphi(\omega)\psi(g)
\end{equation}
To check if the factorization of Eq. (\ref{complete}) is correct we have to verify if it exists a function of $\omega$ such that the impact surface divided by this function turns out to be independent on $\omega$. This is equivalent to say that we would like to observe a complete collapse into a unique curve 
by rescaling with this function the surface $\phi$.
The simplest choice for this scaling function is the average impact function $\langle\phi(\omega, g)\rangle_g\sim\omega^{0.59}$. We report the rescaled functions for different values of $\omega$ in Fig. \ref{collapse} and we observe a nearly perfect collapse for $\omega<30$ within error bars. 
\\
In principle we cannot obtain a perfect factorization because of the dependence of the exponent $\beta$ on $g$. 
Nevertheless we observe that this dependence is very weak, especially for $g>1$ when
$\beta\approx0.60$. In fact in Fig.~\ref{collapse} one can appreciate
some discrepancies only for $g<1$.
We can conclude that the factorization proposed in Eq. (\ref{complete}) approximately holds at least for small values of $\omega$. We also expect that $\varphi(\omega,\Delta t\rightarrow0)$ can be factorized with respect to its variables, instead the situation may be very complicated for the function $\psi(g,\Delta t\rightarrow0)$ since $g$ depend on time. We will consider these points in future investigations.

\section{Conclusions}\label{sec:6}

In this paper we have introduced a model of order book to study the generalized Price Impact Function (Price Impact Surface) and its dependence not only on order volume but also on granularity.
We find that the granularity operates as a strong amplifier of price variations 
when a liquidity crisis takes place.
In particular the dependence of the Price Impact Surface on the granularity is a power law
with exponent nearly equal to $-1$. This result implies that the system response to an incoming order
diverges in the limit of vanishing granularity.
It would be interesting to compare the Price Impact Surface found in the model with the empirical one. 
\\ 
Furthermore agent-based models for financial markets usually do not take into account the problem of finite liquidity. 
In fact a common way to model the price movements in an agent based model is through the Walras' mechanism in which the price adjustments are proportional to the excess demand i.e. $\Delta p=\chi\, ED$. 
The coefficient $\chi$ and the excess demand $ED$ are generally assumed to be independent on granularity. 
If, in first approximation, one interprets the Market Impact Function as the quantity the price movements are proportional to in the Walras' mechanism, then the previous coefficient $\chi$ must depends on granularity. 
Moreover we have observed that the Price Impact Surface can be factorized and we can try to identify the excess demand with $\varphi(\omega)$ and $\chi$ with $\psi(g)$. 
Consequently we aim at introducing this dependence on $g$ in the \textit{workable} agent based model we have introduced in previous papers~\cite{paperoNP,paperoI,paperoII,paperoSOI}. 
In this framework we expect that even small unbalances of the market can produce large price adjustments if a liquidity crisis is present. Therefore the role of the amplification introduced by $\psi(g)$ could be one of the explanation of the breaking of the cause-effect relation and so that small perturbations can produce large fluctuations.

\section*{Acknowledgments}
We are grateful to Doyne Farmer, Gabriele La Spada, Marcello Mezzedimi and the anonymous referee for interesting discussions and suggestions.


\end{document}